\documentclass[12pt]{article}
\usepackage{amssymb}

\newtheorem{theorem}{Theorem}

\newtheorem{lemma}[theorem]{Lemma}

\newtheorem{cor}[theorem]{Corollary}

\newcommand{\eqa}{\begin{eqnarray}}
\newcommand{\eeqa}{\end{eqnarray}}
\newcommand{\beq}{\begin{equation}}
\newcommand{\eeq}{\end{equation}}
\newcommand{\nn}{\nonumber}

\newcommand{\pal}{\partial}

\newcommand{\al}{\alpha}

\newcommand{\lm}{\lambda}

\newcommand{\ve}{\epsilon}

\newcommand{\pf}{\noindent{\it Proof \ }}

\newcommand{\res}{{\rm res}}
\newcommand{\lgl}{\log{L}}
\newcommand{\ld}{\Lambda}
\newcommand{\epf}{$\quad$\hfill
\raisebox{0.11truecm}{\fbox{}}\par\vskip0.4truecm}

\begin{document}

\title {\LARGE The Extended Toda  Hierarchy}
\author{ {Guido Carlet${}^*$ \ \ Boris Dubrovin${}^*$ \ \ Youjin
Zhang${}^{**}$}\\
{\small ${}^{*}$SISSA, Via Beirut 2--4, 34014 Trieste, Italy}\\
{\small ${}^{**}$Department of Mathematical Sciences,
Tsinghua University, Beijing 100084, P.R.China
}}
\maketitle
\begin{abstract}
We present the Lax pair formalism for certain extension of the continuous limit of the
classical Toda lattice hierarchy,
provide a well defined notion of tau function for its solutions, and give an explicit formulation
of the relationship between the $CP^1$ topological sigma model and the extended Toda hierarchy. We also establish an equivalence of the
latter
with certain extension of the nonlinear Schr\"odinger hierarchy.
\end{abstract}

\setcounter{equation}{0}
\setcounter{theorem}{0}

\section{Introduction}

The Toda lattice equation \cite{Toda}
\beq\label{TLE}
\ddot q_{n}= e^{q_{n-1}-q_n} - e^{q_n - q_{n+1}}, \quad -\infty < \, n\, < \infty
\eeq
is one of the prototypical integrable systems
that plays significant role in classical and quantum field theory.
The Toda lattice hierarchy consists of infinitely many evolutionary
differential-difference
equations commuting with (\ref{TLE}).
In this paper
we study this hierarchy from the point of view of 2D topological field theory.
One of the first lessons of this approach \cite{DW, Witten1} is that, one is to
replace the discrete variable $n$ by a continuous one. The result of such
``interpolation'' is the following equation for the function
$q=q(x,t)$
\beq\label{toda1}
\ve^2 q_{tt} = e^{q(x-\ve) -q(x)} -e^{q(x) - q(x+\ve)}.
\eeq
In this equation the cosmological constant plays the role of the
independent variable $x$, the formal small parameter $\ve$ is called the string coupling
constant. Similar interpolation can be applied to the whole Toda lattice
hierarchy.
It is conjectured
that the partition function of the $CP^1$ topological sigma model as the function of the coupling constants of the theory is the
tau function of a particular solution of certain extension of the interpolated
Toda lattice hierarchy. Under such identification the coupling
constant corresponding to the identity primary field $\phi_1\in H^0(CP^1)$ serves as the spatial variable and
that of the remaining primary $\phi_2 \in H^2(CP^1)$ and of the gravitational descendent fields correspond to the time
variables of the hierarchy. Such an extension of the Toda lattice hierarchy is formulated independently
in \cite{getzler, Z}, and the above conjecture is known to be true up to genus one approximation
\cite{D1, DZ1, Egu1, Egu5, Egu6, Z}.
The extended Toda lattice hierarchy is formulated in \cite{getzler, Z} by using the bihamiltonian
structure of the original Toda lattice hierarchy, and is defined by the bihamiltonian
recursion relation. We call this hierarchy {\it the extended Toda hierarchy} in this paper.

To an expert in the theory of integrable systems that might be less motivated by
the eventual applications of the Toda hierarchy to the theory of Gromov - Witten invariants, the importance
of considering the extended Toda hierarchy can also be explained by means of the following argument.
The flows of the usual Toda hierarchy form a complete family, i.e. they span the space of vector fields
commuting with (\ref{TLE}). This fails to be true for the interpolated Toda lattice hierarchy. Indeed, already the spatial
translations $x\mapsto x+c$ do not belong to the linear span of the Toda lattice flows.
One can show,
using the technique of \cite{DZ3} that the flows of extended Toda lattice hierarchy form a complete family of flows
commuting with (\ref{toda1}).

Two important aspects of the theory of extended Toda hierarchy remained unclear,
after \cite{getzler, Z}. The missing points
were the
Lax pair formalism and a well defined notion of tau function for an arbitrary
solution of the extended hierarchy. A
Lax pair formalism is crucial both for the theory of integrability of the hierarchy
and for its applications in physics, while a well defined notion of tau function
for solutions of the hierarchy is needed to formulate explicitly the relation between the
extended Toda hierarchy and the $CP^1$ topological sigma model, or equivalently, to
the theory of Gromov-Witten invariants of $CP^1$ and their gravitational
descendents.

We present in Section 2 a Lax pair formalism of the extended Toda hierarchy. By using
this Lax pair formalism we are able to express in Section 3 the densities of the Hamiltonians of the hierarchy
in terms of the Lax operator, and in Section 4 we give the definition of the tau function for solutions of the
hierarchy by using the result of Section 3. In Section 5 we show that the extended Toda hierarchy is equivalent
to certain extension of the nonlinear Schr\"odinger hierarchy. In the last Section we discuss the
relation of the $CP^1$ topological sigma model with the extended Toda hierarchy.

\section{Formulation of the extended Toda hierarchy}

The Toda lattice equation (\ref{TLE}) describes the motion of one-dimensional particles with exponential
interaction of neighbors \cite{Toda}.
A crucial aspect among the integrability properties of this equation is its Lax pair formalism
given by Flaschka in \cite{Fla}. By introducing the new dependent variables
\beq
v_n=-\frac{\pal q_n}{\pal t},\quad u_n=q_{n-1}-q_n,
\eeq
we can rewrite the Toda lattice equation in the form
\beq
\frac{\pal v_n}{\pal t}=e^{u_{n+1}}-e^{u_n}, \quad
\frac{\pal u_n}{\pal t}=v_n-v_{n-1},\quad n\in {\mathbb Z}.
\eeq
Let $\Lambda$ be the shift operator defined by
$$
\Lambda f_n=f_{n+1}
$$
for any function $f$ on the one dimensional infinite lattice. The Lax operator \footnote{We use here not the original Lax operator introduced in
\cite{Fla} but the one of Ueno and Takasaki \cite{UT}}
is defined by
\beq
{\bar L}=\Lambda+v_n+e^{u_n} \Lambda^{-1}
\eeq
and the Toda lattice equation can be recast into the form
\beq\label{th-0}
\frac{\pal{\bar L}}{\pal t}=
[\Lambda+v_n,{\bar L}].
\eeq
Here the square bracket stands for the usual
commutator of two operators. Related to the Toda lattice equation there is an infinite
family of mutually commuting flows of the form
\beq\label{th-p}
\frac{\pal{\bar L}}{\pal t_p}=\frac{1}{(p+1)!}
[\left({\bar L}^{p+1}\right)_+,{\bar L}],\quad p\ge 0, ~~\frac{\pal}{\pal t_q} \frac{\pal{\bar L}}{\pal t_p} = \frac{\pal}{\pal t_p}
\frac{\pal{\bar L}}{\pal t_q}.
\eeq
This family of evolutionary diferential-difference equations is the so-called Toda lattice
hiearchy. Clearly for $p=0$ the equation (\ref{th-p}) coincides with (\ref{th-0}).

We are to define certain extension of the Toda lattice by constructing another infinite family of
evolutionary equations that commute with
each other and with the flows of the original Toda lattice hierarchy. To this end, we first replace the discrete variable $n$ by a continuous
variable $x$. By interpolating we introduce
the dependent variables
$
u(x),\,v(x)
$
such that
\beq
u_n = u(\ve n), \quad v_n=v(\ve n).
\eeq
Here $\ve$ is a formal parameter that can be viewed as the lattice mesh.
We will also use an alternative notation for the dependent variables
\beq\label{index}
w^1:= v, \quad w^2:= u
\eeq
and we will denote $w=(w^1, w^2)$ the two-component vector.
Then the Toda lattice hierarchy for the functions $w^\al(x, t_0, t_1, \dots)$, $\al=1, 2$  can be recast into the form
\beq\label{td-orig}
\ve\frac{\pal{L}}{\pal t^{2,p}}=\frac{1}{(p+1)!}
[\left({L}^{p+1}\right)_+,{L}],\quad p\ge 0.
\eeq
Here the Lax operator $L$ acting on smooth functions on the line is defined by
\beq
L=\Lambda+v(x)+e^{u(x)} \Lambda^{-1}
\eeq
with $\Lambda$ being defined now as the shift operator
$$
\Lambda=e^{\ve\pal_x}
$$
and the time variables $t^{2,p}$ are obtained from $t_p$ by rescaling $t^{2,p}=\ve\,t_p$. We call this hierarchy
the {\em Toda hierarchy}.

Let us denote ${\cal R}$ the ring of formal power series of the form $
\sum_{k\ge 0} f_k \ve^k$, where $f_k$ are
polynomials of the variables $v(x),u(x), e^{\pm u(x)}$ and the $x$-derivatives
of $v, u$. The gradation on ${\cal R}$ is defined by
\beq\label{grad-A}
\deg v^{(m)}=1-m,\ \deg u^{(m)}=-m,\ \deg
e^{u}=2,\ \deg\ve=1, \quad m\ge 0.
\eeq
Here
\beq
v^{(m)} = \pal_x^m v, \quad u^{(m)}=\pal_x^m u.
\eeq

The equations of Toda hierarchy will be considered as ${\cal R}$-valued vector fields. For example,
the interpolated Toda lattice equation has the form
\eqa\label{flow2}
&&
\frac{\pal v}{\pal t^{2,0}}=\frac1{\ve}\left(e^{u(x+\ve)}-e^{u(x)}\right) = \sum_{k\geq 0} \frac{\ve^k}{(k+1)!} \pal_x^{k+1} e^{u}
\nn\\
&&
\frac{\pal u}{\pal t^{2,0}}=\frac1{\ve}\left(v(x)-v(x-\ve)\right)=\sum_{k\geq 0} (-1)^{k+1} \frac{\ve^k}{(k+1)!} \pal_x^{k+1} v.
\eeqa
Following \cite{DZ3}, we will treat equations of this class as infinite order evolutionary PDEs. For the sake of brevity they
will also be called PDEs in subsequent considerations. The solutions of such PDEs will be considered in the class of formal power series in
$\ve$.

The dressing operators $P$ and $Q$  (see \cite{UT})
\beq\label{PQ-2}
P=\sum_{k\ge 0} p_k \ld^{-k},\quad Q=\sum_{k\ge 0} q_k
\ld^{k},\quad p_0=1
\eeq
can be formally defined by the following identities in the ring of Laurent series in $\Lambda^{-1}$ and $\Lambda$
respectively:
\beq\label{PQ-1}
L=P\Lambda P^{-1}=Q\Lambda^{-1} Q^{-1}.
\eeq
Note that the coefficients $p_k$ and $q_k$ of the dressing operators do not belong to the ring ${\cal R}$ but to a certain extension of it
(see \cite{UT}). The dressing operators are defined up to the
multiplication from the right by operators of the form
$1+\sum_{k\ge 1} \hat p_k \ld^{-k}$ and $\sum_{k\ge 0} \hat q_k
\ld^k$ respectively, where $\hat p_k, \hat q_k$ are some constants.

To construct an extension of the Toda hierarchy we need to introduce the following notion
of the logarithm of the Lax operator $L$:
\beq\label{def-log}
\lgl:=\frac12 \left(P\ve\pal_x P^{-1}-Q\ve \pal_x Q^{-1}\right).
\eeq
Remarkably the above ambiguity in the choice of dressing operators is cancelled in the definition of the
operator $\lgl$. Moreover,
the coefficients of the operator $\lgl$ do belong to ${\cal R}$ as the following
theorem guarantees:

\begin{theorem}\label{log-polyn}
The operator $\lgl$ has the following expression
\beq\label{log-coef-w}
\lgl=\sum_{k\in {\mathbb Z}} g_k \ld^k,\quad g_k=g_k(w,w_x,w_{xx},\dots; \ve)\in {\cal R}.
\eeq
\end{theorem}
\pf Let us first consider the operator $\ve P_x P^{-1}$ where
$P_x=\sum_{k\ge 1} p_{k,x} \ld^{-k}$.
It has the expression
\beq\label{pxp-1}
\ve P_x P^{-1}=\sum_{k\ge 1} a_k \ld^{-k}.
\eeq
From the definition of the dressing operator $P$ it follows that
\beq\label{eq-PP}
[\ve P_x P^{-1}, L^m]=\ve \pal_x L^m,\quad m\ge 1.
\eeq
For any operator $A=\sum Y_k \ld^k$ let us define its residue by
\beq\label{def-res}
\res\, A=Y_0.
\eeq
By taking residue on both sides of (\ref{eq-PP}) with $m=1$ we get
\beq\label{a0-1} a_1(x+\ve)-a_1(x)=-\ve\,\pal_x v(x),
\eeq
which shows that $a_1\in {\cal R}$ (see the explicit formula (\ref{bern}) below). By induction on the index of $a_k$ and by taking
residue on both sides of (\ref{eq-PP}) for general $m$ we show that $a_k\in {\cal R}$ for
$k\ge 1$.

To finish the proof of the theorem, we need to obtain similar result for the operator $Q_x Q^{-1}$.
From (\ref{PQ-2}), (\ref{PQ-1}) we know that the function $q_0$ that appears in the expression
of $Q$ satisfies the relations
\beq\label{A-1}
\frac{q_0(x)}{q_0(x-\ve)}=e^{u(x)}, \quad
\frac{q_{0,x}(x)}{q_0(x)}-\frac{q_{0,x}(x-\ve)}{q_0(x-\ve)}=u_x,
\eeq
from which it follows that
the coefficients of the operator $\tilde L=q_0^{-1} L q_0$ belong to ${\cal R}$. Denote
$
\tilde Q=q_0^{-1} Q.
$
We have by definition the relation
\beq\label{tL-tQ}
\tilde L=\tilde Q \ld^{-1} \tilde Q^{-1}.
\eeq
By using the identity (\ref{tL-tQ}) we can
show, as we did for the operator $\ve P_x P^{-1}$ above, that the
operator $\ve\,{\tilde Q}_x {\tilde Q}^{-1}$ has the expression
$$
\ve\,{\tilde Q}_x {\tilde Q}^{-1}=\sum_{k\ge 1} b_k \ld^{k},\quad
b_k\in {\cal R}.
$$
Then the theorem follows from (\ref{A-1}) and the identities
$$
\lgl=\frac12 \left(\ve\,Q_x Q^{-1}-\ve P_x P^{-1}\right),\quad Q=q_0 {\tilde Q}.
$$
The Theorem is proved.
\epf

The proof of the above theorem also gives an algorithm
of computing the coefficients $g_k$ of the operator $\lgl$
expanded in the form (\ref{log-coef-w}). Indeed, from
(\ref{a0-1}) we obtain
\beq\label{bern}
 a_1(x)=-\sum_{k\ge 0} \frac{B_k}{k!}
(\ve \pal_x)^k v(x)+c
\eeq where the coefficients $B_k$ are the
Bernoulli numbers and $c$ is an integration constant. If we set
$v=e^{u}=0$ in the Lax operator $L$, then the coefficients of
the dressing operator $P$ must be constants, and in this situation
$P_x P^{-1}=0$. This fact implies that the integration constant
$c$ must be equal to zero. Now, if we already obtained the
expression for the first $n-1$ coefficients of $\ve P_x P^{-1}$ that
is expanded in the form (\ref{pxp-1}), then the identity
$$
\res \left([\ve P_x P^{-1}, L^{n}]-\ve \pal_x L^{n}\right)=0
$$
can be written in the form
$$
a_{n}(x+n\ve)-a_{n}(x)=\ve \pal_x W
$$
for some $W\in {\cal R}$.
So we have
$$
a_n(x)=\sum_{k\ge 0} \frac{B_k}{k!} (n \ve \pal_x)^k W.
$$
Here the integration constants also disappear due to the same reason as for the
vanishing of the integration constant $c$ for $a_1$. The
coefficients of the operator $\tilde Q_x \tilde Q^{-1}$ can be
computed in a similar way.

\vskip 0.4truecm
\noindent {\bf Definition.}\ The {\em extended Toda hierarchy} consists of the
evolutionary PDEs that are represented in the following Lax pair formalism:
\beq\label{td-flow-Lax}
\ve \frac{\pal L}{\pal t^{\beta,q}}=[A_{\beta,q},L] :=A_{\beta,q} L-L
A_{\beta,q},\quad \beta=1,2;\ q\ge 0.
\eeq
Here the operators $A_{\beta,q}$  are defined by
\beq\label{def-a-1}
A_{1,q}=\frac2{q!}\left[L^q (\lgl-c_q)\right]_+,\quad
A_{2,q}=\frac1{(q+1)!}\left[L^{q+1}\right]_+,
\eeq
and for any operator $B=\sum B_k \ld^k$, the operator $B_+$ is given by $\sum_{k\ge 0} B_k
\ld^k$. Here the constants $c_q$ are defined as follows
\beq\label{c-q}
c_0=0,  \quad
c_q=1+\frac12+\dots+\frac1{q}.
\eeq

\vskip 0.4truecm

The flows $\frac{\pal}{\pal t^{2,p}},\ p\ge 0$ form the original Toda hierarchy (\ref{td-orig}).
We will see in the next Section that the flows $\frac{\pal}{\pal t^{1,p}},\ p\ge 0$ coincide
with those defined in \cite{getzler,  Z} by using a bihamiltonian recursion relation. In the
literature the explicit Lax pair formalism for these flows exists only for their dispersionless limit
\cite{Egu1, Egu5, Egu6}. To have a more concrete feeling of the form of these flows,
let us write down the first three of them. By the definition
(\ref{def-a-1}), we have
$$
A_{1,0}=\left(P\ve \pal_x P^{-1}-Q\ve \pal_x Q^{-1}\right)_+=\ve Q_x Q^{-1}
=\ve \pal_x-\ve Q\pal_x Q^{-1}.
$$
Since $[Q\ve \pal_x Q^{-1}, L]=0$, we obtain
\beq
\frac{\pal w^\al}{\pal t^{1,0}}=w^\al_x,\quad \al=1,2.
\eeq
So this first flow is just the translation along the spatial variable $x$. The second flow is
the interpolated Toda lattice  equation (\ref{flow2}).

Introduce the following two operators that act on the space of smooth functions of $x$:
\eqa
&&{\cal B}_+ f(x):=(\Lambda-1)^{-1}\ve \pal_x f(x)=\sum_{k\ge 0} \frac{B_k}{k!}(\ve \pal_x)^k f(x),\nn\\
&&
{\cal B}_- f(x):=(1-\Lambda^{-1})^{-1}\ve \pal_x f(x)=\sum_{k\ge 0} \frac{B_k}{k!}(-\ve \pal_x)^k f(x).\label{def-calB}
\eeqa
Here $B_k$ are the Bernoulli numbers. The following operator
\beq
A_{1,1}= \left(\Lambda + v\right) (\ve \pal_x -1) +{\cal B}_+ v(x+\ve)
+ e^u \left[ \ve \pal_x + 1 -{\cal B}_- u(x-\ve)
\right] \Lambda^{-1}
\label{a11-ad}
\eeq
(it differs from the one given by (\ref{def-a-1}) by the operator $-L (1+Q\ve\pal_x Q^{-1})$ commuting with $L$)
gives the Lax representation for the $t^{1,1}$-flow
\eqa
&& \frac{\pal v}{\pal t^{1,1}}=v v_x+\frac1{\ve}\left[e^{u(x+\ve)}\left({\cal B}_- u(x+\ve)-2\right)
-e^{u(x)}\left({\cal B}_- u(x-\ve)-2\right)\right],\nn\\
&&\frac{\pal u}{\pal t^{1,1}}=\frac1{\ve} \left[v(x) \left({\cal B}_- u(x)-2\right)-v(x-\ve)
\left({\cal B}_- u(x-\ve)-2\right)\right.\nn\\
&&\quad \quad \quad \quad\quad \left.+{\cal B}_+ v(x+\ve)-{\cal B}_+ v(x-\ve)\right].
\eeqa

We finish this Section with the following simple statement.

\begin{theorem}
The components of the vector fields of the extended Toda hierarchy
are homogeneous elements of the graded ring ${\cal R}$, of the degree
\beq\label{dg-flow}
\deg\frac{\pal w^\al}{\pal t^{\beta,q}}=q+\mu_\beta-\mu_\al,\quad \al, \beta=1,2,\ q\ge 0.
\eeq
Here $\mu_1=-\frac12, \mu_2=\frac12$.
\end{theorem}

We leave the proof as an exercise for the reader.

\setcounter{equation}{0} \setcounter{theorem}{0}

\section {Bihamiltonian structure of the extended Toda hierarchy}\label{sect-2}
The existence of a bihamiltonian structure for the original Toda lattice hierarchy is well known
(see for example \cite{DZ1, kup-1}). In this Section we are to adopt it to the extended
Toda hierarchy (\ref{td-flow-Lax}). The bihamiltonian structure for the
original Toda hierarchy is given by the following two compatible Poisson brackets
\eqa
&&\{v(x),v(y)\}_1=\{u(x),u(y)\}_1=0,\nn\\
&&\{v(x),u(y)\}_1=\frac{1}{\ve} \left[e^{\epsilon\,\pal_x}-1
\right]\delta(x-y),\label{toda-pb1}\\
&& \{v(x), v(y)\}_2={1\over\epsilon}\left[e^{\epsilon\,\pal_x}
e^{u(x)}-
e^{u(x)} e^{-\epsilon\,\pal_x}\right] \delta(x-y),\nn\\
&& \{ v(x), u(y)\}_2 = {1\over \epsilon}
v(x)\left[e^{\epsilon\,\pal_x}-1 \right]
\delta(x-y)\label{toda-pb2}\\
&& \{ u(x), u(y)\}_2 = {1\over \epsilon} \left[
e^{\epsilon\,\pal_x}-e^{-\epsilon\,\pal_x}\right]\delta(x-y) \nn
\eeqa
In particular, for a local Hamiltonian
$$
H= \int h(w; w_x, w_{xx}, \dots; \ve)\, dx, \quad h(w; w_x, w_{xx}, \dots; \ve)\in {\cal R}
$$
the Hamiltonian system w.r.t. the first Poisson bracket reads
\eqa\label{explicit1}
&&
u_t =\{ u(x), H\}_1 = \frac1{\ve} \left[ 1-e^{-\ve \pal_x}\right] \frac{\delta H}{\delta v(x)}
\nn\\
&&
v_t =\{v(x), H\}_1 = \frac1{\ve} \left[ e^{\ve \pal_x}-1\right] \frac{\delta H}{\delta u(x)}.
\eeqa
The same Hamiltonian will generate a different PDE when the second Poisson bracket is used:
\eqa\label{explicit2}
&&
u_s=\{ u(x), H\}_2 = \frac1{\ve} \left[ \Lambda-\Lambda^{-1}\right] \frac{\delta H}{\delta u(x)} + \frac1{\ve}  \left[ 1-\Lambda^{-1}\right]
\frac{\delta H}{\delta v(x)}
\nn\\
&&
v_s=\{ v(x), H\}_2 =\frac1{\ve} v(x) \left[ \Lambda -1\right] \frac{\delta H}{\delta u(x)} + \frac1{\ve} \left[ \Lambda e^{u(x)} - e^{u(x)}
\Lambda^{-1}\right] \frac{\delta H}{\delta v(x)}.
\eeqa
Here $s$ is the new time variable.

We have the following main theorem of this Section:
\begin{theorem}
The flows of the extended Toda hierarchy (\ref{td-flow-Lax}) are Hamiltonian systems
of the form
\beq
\frac{\pal w^\al}{\pal t^{\beta,q}}=\{w^\al(x),H_{\beta,q}\}_1,\quad \al,\beta=1,2;\ q\ge 0.
\label{td-ham}
\eeq
They satisfy the following bihamiltonian recursion relation
\beq
\{w^\al(x),H_{\beta,q-1}\}_2=(q+\mu_\beta+\frac12)
\{w^\al(x),H_{\beta,q}\}_1+R^\gamma_\beta \{w^\al(x),H_{\gamma,q-1}\}_1.\label{td-recur}
\eeq
Here the Hamiltonians have the form
\beq
H_{\beta,q}=\int h_{\beta,q}(w; w_x, \dots; \epsilon) dx,\quad \beta=1,2; \ q\ge -1
\eeq
with the Hamiltonian densities $h_{\beta,q}=h_{\beta,q}(w; w_x, \dots; \epsilon)\in {\cal R}$ given by
\beq\label{def-h}
h_{1,q}=\frac2{(q+1)!}\,\res\left[ L^{q+1}
(\lgl-c_{q+1})\right],\quad h_{2,q}=\frac1{(q+2)!}\res \, L^{q+2},
\eeq
and
\beq\label{def-mu-R}
R^\gamma_\beta=2 \delta^\gamma_2 \delta_{\beta,1}.
\eeq
\end{theorem}
\pf We first prove that the flows $\frac{\pal}{\pal t^{2,q}}$ have the Hamiltonian form
(\ref{td-ham}). For a 1-form $\sum f_{\al,m}(u,u_x,\dots) du^{\al,m}$ on the jet space, we say that it is
equivalent to zero if it is the $x$-derivative of
another 1-form. We denote this equivalence relation by $\sim$.
Here $w^{\al,m}=\frac{\pal w^\al(x)}{\pal x^m}$. Alternatively,
$$
w^{1,m}= v^{(m)}, \quad w^{2,m} = u^{(m)}.
$$
For example, we have
$$
e^{u} d v_x+e^{u} u_x d v=\pal_x (e^{u} d v)\sim 0.
$$
Under this notation, we can easily verify that
\beq\label{equiv-r}
d h_{2,q}=\sum \frac{\pal h_{2,q}}{\pal u^{\al,m}}\, d u^{\al,m}=\frac1{(q+2)!}\, d\, \res\,
L^{q+2} \sim \frac1{(q+1)!} \res\, L^{q+1} d L
\eeq
where $d L=d v(x)+e^{u(x)} du(x)\,\Lambda^{-1}$.
Expand the operators $A_{\beta,q}$ that are defined in
(\ref{def-a-1}) into the form
\beq\label{expand-A}
A_{1,q}=\sum_{k\ge 0} a_{1,q;k}\, \ld^k,\quad A_{2,q}=\sum_{k\ge 0} a_{2,q;k}\, \ld^{k},
\eeq
then by using the definition of the Hamiltonians $H_{2,q}$ and the equivalence relation (\ref{equiv-r})
we deduce the validity of the following identities:
\beq\label{dH2-u12}
\frac{\delta H_{2,q}}{\delta
v}=a_{2,q;0}(x),\quad \frac{\delta H_{2,q}}{\delta
u}=a_{2,q;1}(x-\ve) e^{u(x)}.
\eeq
So from the definition of the first Poisson bracket we have
\eqa
&& \{v(x),H_{2,q}\}_1=\frac1{\ve}\left(a_{2,q,1}(x) e^{u(x+\ve)}-a_{2,q,1}(x-\ve)
e^{u(x)}\right).
\nn\\
&&\{u(x), H_{2,q}\}_1=\frac1{\ve}\left(a_{2,q,0}(x)-a_{2,q,0}(x-\ve)\right)
\eeqa
which yields the Hamiltonian form (\ref{td-ham}) of the flows $\frac{\pal}{\pal t^{2,q}}$.

To prove that the flows $\frac{\pal}{\pal t^{1,q}}$ are also
Hamiltonian systems with respect to the first Poisson bracket,
we need first to show the validity of the following equivalence relation:
\beq\label{dlgl}
\res\left(L^q\, d \lgl\right) \sim \res \left(L^{q-1} d L\right).
\eeq
Indeed, from the commutativity of the operators
$L$ and $P \ve\pal_x P^{-1}$ we obtain
\eqa
&&d\,\res \left[L^{q}
e^{P \ve\pal_x P^{-1}}\right]\nn\\
&& \sim q \res \left[\,L^{q-1}
e^{P \ve\pal_x P^{-1}} d L\right]+\res \left[L^{q} \sum_{k\ge 1} \frac1{(k-1)!}
(P\ve\pal_x P^{-1})^{k-1} d (P \ve\pal_x P^{-1})\right]\nn\\
&& =q \,\res L^{q} d L+\res \left[ L^{q} e^{P\ve\pal_x P^{-1}} d
(P \ve\pal_x P^{-1})\right]. \nn
\eeqa
So from the obvious relations
$$
L^{q} e^{P \ve\pal_x P^{-1}}=L^{q+1},\quad
d\, \res L^{q+1}\sim (q+1) \res L^q d L
$$
we arrive at
\beq\label{dlgl-2} \res \left[L^q d (P\ve \pal_x
P^{-1})\right] \sim \res L^{q-1} d L.
\eeq
In a similar way we obtain the following equivalence relation
\beq\label{dlgl-3}
\res \left[L^q d (Q\ve\pal_x Q^{-1})\right]\sim -\res L^{q-1} d L.
\eeq
The equivalence relation (\ref{dlgl}) now readily follows from the above two equations.
By using (\ref{dlgl}) we obtain
\eqa
&&d h_{1,q}=\frac2{(q+1)!}\,d\,\res \left[L^{q+1}
\left(\lgl-c_{q+1}\right) \right]
\nn\\
&& \sim \frac2{q!}\,\res\left[L^q
\left(\lgl-c_{q+1}\right) d L\right]+ \frac2{(q+1)!}\,\res \left[L^q d L\right]\nn\\
&&=\frac2{q!}\,\res\left[L^q \left(\lgl-c_{q}\right) d L\right].
\eeqa
It yields the following identities
\beq\label{dH1-u12}
\frac{\delta H_{1,q}}{\delta v}=a_{1,q;0}(x),\quad \frac{\delta H_{1,q}}
{\delta u}=a_{1,q;1}(x-\ve) e^{u(x)}.
\eeq
Here $a_{\al,p;k}$ are defined in (\ref{expand-A}). From the above identities we see that
the flows $\frac{\pal}{\pal t^{1,q}}$ that is defined by (\ref{td-flow-Lax}) are Hamiltonian systems
of the form (\ref{td-ham}).

We now proceed to proving the bihamiltonian recursion relation (\ref{td-recur}). In the case of
$\al=1,\beta=2$, we can rewrite (\ref{td-recur}) by using the identities (\ref{dH2-u12}) into the form
\eqa
&&\left[\Lambda e^{u(x)}-e^{u(x)} \Lambda^{-1}\right] a_{2,q-1;0}(x)+
v(x) \left[\Lambda-1\right] a_{2,q-1;1}(x-\ve) e^{u(x)}\nn\\
&&
=(q+1)\left[a_{2,q;1}(x) e^{u(x+\ve)}-a_{2,q;1}(x-\ve) e^{u(x)}\right].\label{pre-recur}
\eeqa
On the other hand, from the first and the second equality of the relation
\beq\label{pre-b}
(q+1)\frac{1}{(q+1)!}L^{q+1}=L\,\frac1{q!} L^{q}=\frac1{q!} L^{q} L
\eeq
we obtain respectively the following identities
\eqa
&&(q+1) a_{2,q;1}(x)=a_{2,q-1;0}(x+\ve)+v(x) a_{2,q-1;1}(x)+e^{u(x)} a_{2,q-1;2}(x-\ve),\nn\\
&&(q+1) a_{2,q;1}(x)=a_{2,q-1;0}(x)+v(x+\ve) a_{2,q-1;1}(x)+e^{u(x+2\ve)} a_{2,q-1;2}(x).\nn
\eeqa
The recursion relation (\ref{pre-recur}) can be easily verified by substituting
the above two expressions of $a_{2,q;1}(x)$ into its right hand side.
In the case of $\al=2,\beta=2$, the recursion relation (\ref{td-recur}) can be also verified by using the
identities in (\ref{pre-b}). Finally, for the case of $\beta=1$ the recursion relation (\ref{td-recur})
follows from the following trivial identities
\eqa
&&q\, \frac{2}{q!} L^{q} \left(\lgl-c_{q}\right)=L\,
\frac{2}{(q-1)!}
L^{q-1} \left(\lgl-c_{q-1}\right)-2\,\frac1{q!} L^q\nn\\
&&=\frac{2}{(q-1)!} L^{q-1} \left(\lgl-c_{q-1}\right)\,
L-2\,\frac1{q!} L^q.\nn
\eeqa
Theorem is proved. \epf

In \cite{getzler,  Z} an extended Toda hierarchy was defined by using the bihamiltonian recursion
relation (\ref{td-recur}), and the Hamiltonians are defined implicitly from this recursion relation.
The above theorem shows that this extended Toda hierarchy coincides with the one that is defined
by (\ref{td-flow-Lax}), it also gives an explicit expression of the densities of the Hamiltonians
of the hierarchy. We list here the first few of them
\eqa
&&h_{1,-1}={\cal B}_{-} u(x),\quad h_{2,-1}=v(x),\nn\\
&&h_{1,0}={\cal B}_+(v(x)+v(x+\ve))-2\, v(x)+v(x)\,{\cal B}_{-} u(x),\nn\\
&&h_{2,0}=v(x)^2+e^{u(x)}+e^{u(x+\ve)},
\eeqa
where the operators ${\cal B}_{\pm}$ are defined in (\ref{def-calB}).
We will see below that these densities of the Hamiltonians possess an important
symmetry property which will be used to define the tau functions for solutions of the extended
Toda hierarchy.

\setcounter{equation}{0} \setcounter{theorem}{0}
\section{Tau functions for the extended Toda hierarchy}
We now proceed to define the tau functions for solutions of the extended Toda hierarchy.
Denote by ${\tilde {\cal R}}$ the subset of homogeneous elements of the ring ${\cal R}$, i.e.
elements of the form
$$
f=\sum_{k\ge 0} f_k(u,u_x,\dots) \ve^k
$$
where $f_k$ are homogeneous polynomials of $e^{\pm
u}$, $v^{(m)}$, $u^{(m)}$ for $ m\ge 0$ with $\deg f_k=\deg f-k$. From the definition of the extended Toda
hierarchy (\ref{td-flow-Lax}) and the densities of the hamiltonian (\ref{def-h}) we know that
$
\frac{\pal w^\al}{\pal t^{\beta,q}},\,
h_{\beta,q}\in{\tilde {\cal R}}.
$
The degrees of the flows are given in (\ref{dg-flow}) and the degrees of $h_{\beta,q}$ are given by
$$
\deg h_{\beta,q}=q+\frac32+\mu_\beta.
$$

\begin{lemma}
The following formulae hold true:
\beq\label{dt-lgl}
\frac{\pal \lgl}{\pal t^{\beta,q}}=[A_{\beta,q},\lgl].,\quad \beta=1,2,\ q\ge 0.
\eeq
\end{lemma}
\pf
From (\ref{eq-PP}) we have
$$
[\frac{\pal (P\ve\pal_x P^{-1})}{\pal
t^{\beta,q}},L^m]+[P\ve\pal_x P^{-1}, [A_{\beta,q},L^m]]=0.
$$
The Jacobi identity and the commutativity between the
operators $L$ and $P\ve \pal_x P^{-1}$ then imply the following identity
$$
[\frac{\pal (P\ve\pal_x P^{-1})}{\pal t^{\beta,q}}-[A_{\beta,q},
P\ve\pal_x P^{-1}],L^m]=0.
$$
Since the operator $\frac{\pal (P\ve\pal_x P^{-1})}{\pal
t^{\beta,q}}-[A_{\beta,q}, P\ve\pal_x P^{-1}]$
has the form $\sum_{k\ge 1} f_k \ld^{-k}$ with coefficients $f_k$ being elements of
$\tilde {\cal R}$, we obtain from the last equality the formula
$$
\frac{\pal (P\ve\pal_x P^{-1})}{\pal t^{\beta,q}}-[A_{\beta,q},
P\ve\pal_x P^{-1}]=0.
$$
In a similarly we can also get the formula
$$
\frac{\pal (Q\ve\pal_x Q^{-1})}{\pal t^{\beta,q}}-[A_{\beta,q},
Q\ve\pal_x Q^{-1}]=0.
$$
So the lemma follows from the definition of $\lgl$ and from the last two identities.
\epf

We introduce now the functions $\Omega_{\al,p;\beta,q}$ by the formula
\beq\label{omega-def-1} \frac1{\ve}\,
(\ld-1)\Omega_{\al,p;\beta,q}:=\frac{\pal h_{\al,p-1}}{\pal
t^{\beta,q}} =\left\{\begin{array}{ll} \frac2{p!}\,
\res\left(\left[A_{\beta,q},L^{p}
(\lgl-c_{p})\right]\right),& \al=1;\\
\frac1{(p+1)!}\,\res \left[A_{\beta,q},L^{p+1}\right],& \al=2
\end{array}\right.
\eeq
and by the homogeneity condition
\beq\label{omega-def-2}
\Omega_{\al,p;\beta,q}\in {\tilde {\cal R}},\quad
\deg\Omega_{\al,p;\beta,q} =p+q+1+\mu_\al+\mu_\beta,\quad
\al,\beta=1,2,\  p,q\ge 0.
\eeq
Note that in the above definition the second equality of (\ref{omega-def-1}) follows from the
definition (\ref{td-flow-Lax}), (\ref{def-h}) and from the above lemma. The r.h.s. is a total $x$-derivative
of a homogeneous element in ${\cal R}$. Therefore $\Omega_{\al,p; \beta,q}\in \tilde{\cal R}$ and
the conditions (\ref{omega-def-1}) and
(\ref{omega-def-2}) specify $\Omega_{\al,p;\beta,q}$ uniquely. The only exception is $\Omega_{1,0; 1,0}$ that
should be a homogeneous element
of the degree 0. This is set to be
$$
\Omega_{1,0; 1,0}=u.
$$
The
following Theorem shows that $\Omega_{\al,p; \beta,q}$ is symmetric with respect to
the pair of its indices ${(\al,p)}$ and $(\beta,q)$:

\begin{theorem}
The extended Toda hierarchy has the following tau-symmetry property:
\beq
\frac{\pal h_{\al,p-1}}{\pal t^{\beta,q}}=\frac{\pal
h_{\beta,q-1}}{\pal t^{\al,p}},\quad \al,\beta=1,2,\ p,q\ge 0.
\eeq
\end{theorem}
\pf Let us prove the theorem for the case when $\al=1, \beta=2$,
other cases are proved in a similar way. From the second identity
of (\ref{omega-def-1}) we obtain
\eqa
&&\frac{\pal h_{1,p-1}}{\pal t^{2,q}}=\frac2{p!\,(q+1)!}\,\res [(L^{q+1})_+, L^p (\lgl-c_p)]\nn\\
&& =\frac2{p!\,(q+1)!}\,\res [-(L^{q+1})_-, L^p (\lgl-c_p)]\nn\\
&&=\frac2{p!\,(q+1)!}\,\res [(L^p (\lgl-c_p))_+,(L^{q+1})_-]\nn\\
&& =\frac2{p!\,(q+1)!}\,\res [(L^p (\lgl-c_p))_+,L^{q+1}]=\frac{\pal h_{2,q-1}}{\pal t^{1,p}}.
\eeqa
Theorem is proved.\epf

From the above theorem and the definition (\ref{omega-def-1}) it follows that
$\frac{\pal \Omega_{\al,p;\beta,q}}{\pal t^{\sigma,k}} $ is symmetric w.r.t. the three pairs of indices
$(\al,p), (\beta,q), (\sigma,k)$. This property justifies the following definition of
tau function for the extended Toda hierarchy:

\vskip 0.4truecm \noindent{\bf Definition.}\ For any solution of
the extended Toda hierarchy there exists a function $\tau$ of the
spatial and time variables $x, t^{\al,p}, \al=1,2, p\ge 0$ and of $\ve$ such
that
\beq\label{def-omega} \Omega_{\al,p;\beta,q}=\ve^2 \frac{\pal^2\log\tau} {\pal t^{\al,p}\pal t^{\beta,q}}
\eeq
hold true for any $\al,\beta=1,2,\ p,q\ge 0$.
\vskip 0.4truecm

Recall that the solutions considered in this paper are assumed to be formal power series in $\ve$.

Since the first flow $\frac{\pal}{\pal t^{1,0}}$ of the extended
Toda hierarchy coincides with the translation along the spatial variable $x$,
we can modify the above definition of the tau function by requiring that
\beq\label{t10-x}
\frac{\pal\log\tau}{\pal t^{1,0}}=\frac{\pal\log\tau}{\pal x}.
\eeq

\begin{cor} The densities of the Hamiltonians of the extened Toda hierarchy
have the following expressions in terms of the $tau$ function:
\beq
h_{\al,p}=\ve (\ld-1)\frac{\pal\log\tau}{\pal t^{\al,p+1}},\quad \al=1,2,\ p\ge -1.
\eeq
\end{cor}
\pf From the definition of $\Omega_{\al,p;1,0}$ we get
$$
h_{\al,p-1}=\sum_{k\ge 1} \frac{\ve^{k-1}}{k!}\,\pal_x^{k-1}
\Omega_{\al,p;1,0} =\sum_{k\ge
1}\frac{\ve^{k+1}}{k!}\,\pal_x^{k-1} \frac{\pal^2\log \tau} {\pal
t^{\al,p}\pal t^{1,0}}=\ve (\ld-1) \frac{\pal\log\tau}{\pal
t^{\al,p}}.
$$
Here we used (\ref{t10-x}). The corollary is proved.\epf

Our notion of tau function for the extended Toda hierarchy follows that of Date,
Jimbo, Kashiwara and Miwa designed for the KP hierarchy \cite{djkm83}. Note that
the above corollary implies in particular the following relations of the
dependent variables $v, u$ of the extended Toda hierarchy with the tau function:
\beq
v=\ve \frac{\pal}{\pal t^{2,0}}\log \frac{\tau(x+\ve)}{\tau(x)},\quad
u=\log\frac {\tau(x+\ve)\tau(x-\ve)}{\tau^2(x)}.
\eeq
In this formula we omit the dependence of the tau function on all the times $t^{\al,p}$ but the very first one $t^{1,0}=x$.

{\bf Remark 1}. The above formulae mean that, the dependent variables $u$, $v$ of the extended Toda hierarchy are not normal coordinates
in the sense of \cite{DZ3}. Because of this the relationships between the tau-function and the Hamiltonian densities
in the present paper look more complicated than in the general setting of \cite{DZ3}.

If we return back to the variable $q_n$ of the original Toda lattice equation (\ref{TLE}),
then from the above relation we have
\beq
q_n=\log\frac{\tau(n)}{\tau(n-1)}.
\eeq
So the tau function for the extended Toda hierarchy also agrees with the function that was
introduced by Hirota and Satsuma \cite{hirota} to convert the Toda lattice equation into a bilinear form.
We will discuss the bilinear formulation of the extended Toda hierarchy in a separate publication.

{\bf Remark 2}. The dispersionless limit $\ve\to 0$ of the bihamiltonian structure (\ref{toda-pb1}), (\ref{toda-pb2}) coincides with the canonical Poisson pencil
on the loop space ${\cal L}(M)$ of the Frobenius manifold $M=M_{\tilde W^{(1)}(A_1)}$ constructed in \cite{DZ0} on the orbit space of the
extended affine Weyl group $\tilde W^{(1)}(A_1)$. The Frobenius manifolds $M_{\tilde
W^{(k)}(A_{k+m-1})}$
on the orbit spaces of more general extended affine Weyl groups $\tilde
W^{(k)}(A_{k+m-1})$ of the $A$-series are obtained by the dispersionless limits of extended Toda-like systems associated with the difference
Lax
operators of the form
$$
L=\Lambda^k + a_1(x) \Lambda^{k-1} + \dots + a_{k+m}(x) \Lambda^{-m}, \quad a_{k+m}(x)\neq 0.
$$
This extended hierarchy coincides with the one associated with the Frobenius manifold $M_{\tilde
W^{(k)}(A_{k+m-1})}$ according to the general scheme of \cite{DZ3}. We will give details in a separate publication.
Recall that, in \cite{DZ0} there were also constructed Frobenius manifolds on the orbit spaces
of extended affine Weyl groups associated with the Dynkin diagrams of the $B\, C\, D\, E\, F\, G$ series.
At the moment we do not know how to construct Lax representation of the integrable hierarchies
associated, according to the results of \cite{DZ3}, with these Frobenius manifolds. We plan to study this problem
in subsequent publications.

\setcounter{equation}{0} \setcounter{theorem}{0}
\section {An alternative representation of the extended Toda hierarchy---the extended NLS hierarchy}

In this Section we present an alternative representation of the extended Toda hierarchy that is defined
in the Section 2. We are to choose $t^{2,0}$ as spatial variable and write down the evolutionary
PDEs that are satisfied by the functions $u, v$ under this new spatial variable. Let us redenote the
time variables as follows:
\beq
T^{1,p}=t^{2,p},\quad T^{2,p}=t^{1,p},\quad p\ge 0,
\eeq
and specify $X=T^{1,0}$ as the spatial variable. For the convenience of presentation, we use the following
quantities as the dependent variables:
\beq\label{leg1}
\tilde w^1(X,T)\equiv \varphi=v(x-\ve),\quad \tilde w^2(X,T)\equiv \rho=e^{u(x)}.
\eeq
In terms of the tau function of the extended Toda hierarchy, these new dependent variables have
the expression
\beq
\varphi=\ve (1-\ld^{-1})\frac{\pal\log\tau}{\pal T^{1,0}}, \quad
\rho=\exp[(1-\ld^{-1})(\ld-1)\log\tau].
\eeq

Let us first proceed to writing down the Lax pair formalism for the hierarchy that is
satisfied by $\varphi(X,T), \rho(X,T)$. Note that the extended Toda hierarchy is the
compatibility condition of the following linear systems
\eqa
&&L\psi=\lm \psi,\label{eigen-prob}\\
&&\frac{\pal\psi}{\pal t^{\al,p}}=\ve^{-1} A_{\al,p}\psi,\label{KB-LP}
\eeqa
where $A_{\al,p}$ are defined in (\ref{def-a-1}) and $\lambda$ is the spectral parameter.
By using the equation $\ve\,\pal_{t^{2,0}} \psi=(\ld+v) \psi$, we can rewrite the linear system
(\ref{eigen-prob}) in the form
\beq\label{KB-Lax-a}
{\cal L}\,\psi=\lm \psi
\eeq
with the operator ${\cal L}$ defined by
\beq
{\cal L}=\ve \pal_X+\rho (\ve\pal_X-\varphi)^{-1}.
\eeq
Here the pseudo-differential operator $(\ve\pal_X-\varphi)^{-1}$ has the expansion
\beq
(\ve\pal_X-\varphi)^{-1}=\sum_{k\ge 1} a_i (\ve\,\pal_X)^{-k}
\eeq
and the coefficients $a_k$ are uniquely defined by the relation
\beq
(\ve \pal_X-\varphi)(\sum_{k\ge 1} a_i (\ve\,\pal_X)^{-k})=1.
\eeq
For example, we have
$$
a_1=1,\ a_2=\varphi,\ a_3=-\ve \varphi_X+\varphi^2.
$$
We can also reexpress the operators $A_{\al,p}$ as differential operators in $\ve \pal_X$.
This can be easily done by the substitution
\beq\label{sub1}
\ld^k\mapsto (\ve \pal_X-\varphi(x+(k-1)\ve))\dots (\ve\pal_X -\varphi(x)).
\eeq
So the linear systems in (\ref{KB-LP}) can be expressed in form
\beq\label{KB-Lax-b}
\frac{\pal\psi}{\pal T^{\al,p}}=\ve^{-1}\,{\cal A}_{\al,p} \psi
\eeq
with
\beq {\cal A}_{1,p}=\frac1{(p+1)!} \left({\cal L}^{p+1}\right)_+, \quad {\cal A}_{2,p}=\frac1{p!}
\left({\cal L}^{p}(\log{\cal L} -c_p)\right)_+,
\eeq
the subscript $+$ here means to take the differential part of a pseudo-differential
operator. The pseudo-differential operator $\log{\cal L}$ is obtained from $\lgl$ by the
substitution of (\ref{sub1}) and
\beq
\ld^{-k}\mapsto (\ve\pal_X-\varphi(x-(k-1)\ve))^{-1}\dots (\ve\pal_X-\varphi(x))^{-1}.
\eeq
The coefficients of $\log{\cal L}$ can be expressed in terms of the new dependent variables
$\varphi, \rho$ and their $X$-derivatives. This can be achieved by using the system of
equations (\ref{flow2}) to express $\frac{\pal^m \tilde w^\al}{\pal x^m},\ m\ge 1$ in terms of $\varphi, \rho$
and their $X$-derivatives. For example, we have
\eqa\label{T20-a}
&&\frac{\pal \varphi}{\pal x}=\pal_X\left[\log \rho+\frac{\ve^2}{12 \rho^3}
\left(\rho \rho_{XX}-\rho_X^2-\rho \varphi_X^2\right)+{\cal O}(\ve^4)\right],\\
&&\frac{\pal \rho}{\pal x}=\pal_X\left[\varphi-\frac{\ve^2}{6 \rho^2}
\left(\rho \varphi_{XX}-\varphi_X \rho_X\right)+{\cal
O}(\ve^4)\right].\label{T20-b}
\eeqa

Now the compatibility condition of the linear systems (\ref{KB-Lax-a}), (\ref{KB-Lax-b}) takes the
form
\beq\label{KB-Lax}
\ve\,\frac{\pal{\cal L}}{\pal T^{\al,p}}=[{\cal A}_{\al,p},{\cal L}], \quad \al=1,2,\ p\ge 0.
\eeq
The $T^{1,0}$-flow coincides with the shift along $X$, and the $T^{2,0}$-flow
is given by (\ref{T20-a}) and (\ref{T20-b}).
The $T^{1,1}$-flow has the form
\eqa\label{K-B}
&&\frac{\pal \varphi}{\pal T^{1,1}}=\pal_X \left(-\ve\, \varphi_{X}+\varphi^2+2\, \rho\right),\nn\\
&&\frac{\pal \rho}{\pal T^{1,1}}=\pal_X \left(\ve\, \rho_{X}+2\, \varphi \rho\right).
\eeqa
This integrable system appears in the study of nonlinear water waves in \cite{broer,kaup}.
In terms of the new variables
\beq
q=e^{\ve^{-1}\pal_X^{-1} v}=\rho e^{\ve^{-1}\pal_X^{-1} \varphi},
\quad r=e^{u} e^{-\ve^{-1}\pal_X^{-1}
v}=e^{-\ve^{-1}\pal_X^{-1} \varphi},
\eeq
or, equivalently,
$$
\rho=q\, r, \quad \varphi=-\ve \frac{r_X}{r},
$$
the above system takes the form
\beq
\frac{\pal q}{\pal T^{1,1}}=\ve\, q_{XX}+2\ve^{-1} q^2 r,\quad
\frac{\pal r}{\pal T^{1,1}}=-\ve\, r_{XX}-2 \ve^{-1}q r^2.\label{NLS-qr}
\eeq
These functions have the the following simple expressions in terms of the
tau function of the extended Toda hierarchy:
\beq q=\frac{\tau(x+\ve)}{\tau(x)},\quad
r=\frac{\tau(x-\ve)}{\tau(x)}.
\eeq
Under the constraints  $\ve=i, r=\pm q^*$ the system (\ref{NLS-qr}) is reduced
to the well known nonlinear Schr\"odinger equation (NLS) \cite{Z-S}.
Due to this fact, we will call the hierarchy (\ref{KB-Lax}) the {\em extended NLS hierarchy}.

The extended NLS hierarchy also possesses a bihamiltonian structure. The related compatible
Poisson brackets are given by
\eqa
&&\{\varphi(X), \varphi(Y)\}_1=\{\rho(X), \rho(Y)\}_1=0,\nn\\
&&\{\varphi(X), \rho(Y)\}_1=\delta'(X-Y).\\
&&\{\varphi(X), \varphi(Y)\}_2=2 \delta'(X-Y),\nn\\
&&\{\varphi(X), \rho(Y)\}_2=\varphi(X) \delta'(X-Y)+\varphi_X \delta(X-Y)-\ve
\delta''(X-Y),\nn\\
&&\{\rho(X), \rho(Y)\}_2=\left[\rho(X)\pal_X+\pal_X \rho(X)\right] \delta(X-Y).
\eeqa
This Poisson pencil was given in \cite{bonora} for the bihamiltonian structure of the system (\ref{K-B}).
It is easy to verify that the extended NLS hierarchy hierarchy (\ref{KB-Lax}) has the Hamiltonian form
\beq\label{KB-H}
\frac{\pal \tilde w^\al}{\pal T^{\beta,q}}=\{\tilde w^\al(X),
{\tilde H}_{\beta,q}\}_1, \quad \al,\beta=1,2,\ q\ge 0.
\eeq
Here the Hamiltonians ${\tilde H}_{\beta,q}=\int {\tilde h}_{\beta,q}dX$ are defined by
\beq
{\tilde h}_{1,q}=\frac{1}{(q+2)!}\, \res\, {\cal L}^{q+2},\quad {\tilde h}_{2,q}=\frac2{(q+1)!} \res \left[{\cal L}^{q+1}
(\log{\cal L}-c_{q+1})\right]
\eeq
and the residue of a pseudo-differential operator equals the coefficient of
$\pal_X^{-1}$. The hierarchy satisfies the following bihamiltonian recursion relation:
\eqa
&&\{\tilde w^\al(X), {\tilde H}_{\beta,q-1}\}_2=(q+\frac12+{\tilde \mu}_\beta) \{\tilde w^\al(X),
{\tilde H}_{\beta,q}\}_1+{\tilde R}^\gamma_\beta\,
\{\tilde w^\al(X), {\tilde H}_{\gamma,q-1}\}_1,\nn\\
&&\quad \al,\beta=1,2,\ q\ge 0.
\eeqa
Here ${\tilde \mu}_1=-{\tilde \mu}_2=\frac12,\ {\tilde R}^\al_\beta
=2\delta^\al_1 \delta_{\beta 2}$.

{\bf Remark}. In the ``dispersionless'' limit $\ve\to 0$ the substitution (\ref{leg1}) becomes
\beq\label{leg2}
\varphi=v, \quad \rho = e^u.
\eeq
This coincides with the Legendre type transformation $S_2$ of \cite{cime} (see Appendix B) transforming the Frobenius
manifold associated with Toda lattice with the potential
$$
F_{\rm Toda} = \frac12 v^2 u + e^u
$$
to the Frobenius manifold associated with NLS with the potential
$$
F_{\rm NLS} =\frac12 \varphi^2 \rho +\frac12 \rho^2 \left[ \log\rho -\frac32\right]
$$
(see Example B.1 in \cite{cime}). It looks plausible that the
trick similar to the above one will work also for an arbitrary
semisimple Frobenius manifold in order to lift the Legendre-type
trasnforms of the Frobenius manifold to a transformation of the
integrable hierarchy associated with this manifold. We will
describe these transformations in a separate publication.

\section{The extended Toda hierarchy and the $CP^1$ topological sigma model}
Let $\phi_1=1\in H^0(CP^1)$, $\phi_2=\omega\in H^2(CP^1)$ be the two primary fields for the  $CP^1$ topological sigma model.
The 2-form $\omega$ is assumed to be normalized by the condition
$$
\int_{CP^1}\omega=1.
$$
The free energy of the $CP^1$ topological sigma-model is a function of infinite number of {\it coupling parameters}
$$
{\bf t}=(t^{1,0}, t^{2,0}, t^{1,1}, t^{2,1}, \dots)
$$
and of $\ve$
defined by
the following genus expansion form:
\beq\label{free}
{\cal F}({\bf t}; \ve)=\sum_{g\ge 0}\ve^{2g-2} {\cal F}_g({\bf t}).
\eeq
The parameter $\ve$ is called here the string coupling constant, and the function ${\cal F}_g={\cal F}_g({\bf t})$
is called the genus $g$ free energy which is given by
\beq
{\cal F}_g=\sum_{} \frac1{m!} t^{\al_1,p_1}\dots t^{\al_m,p_m} \langle\tau_{p_1}(\phi_{\al_1})\dots
\tau_{p_m}(\phi_{\al_m})\rangle_g,
\eeq
where $\tau_p(\phi_\al)$ are the gravitational descendent of the primary fields with coupling constants $t^{\al,p}$,
and the rational numbers $\langle\tau_{p_1}(\phi_{\al_1})\dots \tau_{p_m}(\phi_{\al_m})\rangle_g$ are given by the genus $g$ Gromov-Witten
invariants and their descendents of $CP^1$:
\beq
\langle\tau_{p_1}(\phi_{\al_1})\dots \tau_{p_m}(\phi_{\al_m})\rangle_g = \sum_{\beta} q^\beta\int_{[\bar M_{g,m}(CP^1,\beta)]^{\rm virt}}
{\rm ev}_1^* \phi_{\al_1} \wedge \psi_1^{p_1} \wedge \dots \wedge {\rm ev}_m^* \phi_{\al_m} \wedge \psi_m^{p_m}.
\eeq
Here $\bar M_{g,m}(CP^1,\beta)$ is the moduli space of stable curves of genus $g$ with $m$ markings of the given degree $\beta\in H_2 (CP^1;
{\mathbb Z})$, ${\rm ev}_i$ is the evaluation map
$$
{\rm ev}_i: \bar M_{g,m}(CP^1,\beta)\to CP^1
$$
corresponding to the $i$-th marking,
$\psi_i$ is the first Chern class of the tautological line bundle over the moduli space corresponding to the $i$-th marking. According to the divisor
axiom \cite{KM} the indeterminate $q$ can be absorbed by shift $t^{2,0}\mapsto t^{2,0}-\log q$; we will assume that such a shift has already
been performed. So the free energy (\ref{free}) does not depend on $q$.

The conjectural relation of the $CP^1$ topological sigma model with the extended Toda hierarchy can now be
stated in a similar way as the Kontsevich-Witten result \cite{Witten1, Kon, Witten2}\footnote{An alternative
proof was given recently by Okounkov in \cite{oko}} does for the relation of the 2d topological gravity
with the KdV hierarchy. Namely,

\begin{theorem} \label{t-c} $[${\bf Toda conjecture}$]$ The functions
\eqa\label{td-conj}
&&
u(x, {\bf t};\ve) = {\cal F}(t^{1,0}+x +\ve) - 2 {\cal F}(t^{1,0}+x) + {\cal F}(t^{1,0}+x-\ve)
\nn\\
&&
v(x, {\bf t}; \ve) = \ve \frac{\pal}{\pal t^{2,0}}\left[ {\cal F}(t^{1,0}+x+\ve) - {\cal F}(t^{1,0}+x)\right]
\eeqa
satisfy the equations of the extended Toda hierarchy (\ref{td-flow-Lax}).
In these formulae we write explicitly down only
those arguments of the function ${\cal F}$ that have been modified.
This particular solution is uniquelly specified by
the string equation
\beq
\sum_{p\ge 1} t^{\al,p}\frac{\pal {\cal F}}{\pal t^{\al,p-1}}+\frac1{\ve^2} t^{1,0}\,t^{2,0}=\frac{\pal{\cal F}}{\pal t^{1,0}}.
\eeq
\end{theorem}

\newcommand{\dla}{\langle\hskip -0.1truecm\langle}
\newcommand{\dra}{\rangle\hskip -0.1truecm\rangle}
The bihamiltonian description of the extended Toda hierarchy obtained in Section 3 above along with the tau-structure described in
Section 4 enables one to rewrite the bihamiltonain recursion (\ref{td-recur}) in the form of a recursion for the correlators of
the $CP^1$ topological sigma-model. Namely, let us introduce, following \cite{Witten1}, the functions
$\dla \tau_p(\phi_\al)\tau_q(\phi_\beta)\dots\dra$ of ${\bf t}$, $\ve$ by
\beq\label{w-corr}
\dla\tau_{p_1}(\phi_{\al_1})\dots \tau_{p_m}(\phi_{\al_m})\dra=\ve^m \frac{\pal}{\pal t^{\al_1, p_1}} \dots \frac{\pal}{\pal t^{\al_m, p_m}}
{\cal F}({\bf t}; \ve).
\eeq
Then the following recursion relations hold true
\eqa
&&
(n+1) (\Lambda -1) \dla\tau_n (\omega)\dra
\nn\\
&&
= v \, (\Lambda-1) \dla\tau_{n-1}(\omega)\dra + (\Lambda+1) \dla\tau_0(\omega) \tau_{n-1}(\omega)\dra,\label{tau-rec1}
\\
&&\nn \\&&
n\, (\Lambda-1) \dla\tau_{n}(1)\dra
\nn\\
&&
= v\, (\Lambda-1)\dla\tau_{n-1}(1)\dra-2(\Lambda-1)\dla\tau_{n-1}(\omega)\dra\nn\\
&&\quad
+(\Lambda+1) \dla\tau_0(\omega)\tau_{n-1}(1)\dra.\label{tau-rec2}
\eeqa
In these recursion relations
$$
\Lambda = \exp{\ve \frac{\pal}{\pal t^{1,0}}}, \quad v = \ve (\Lambda-1) \frac{\pal {\cal F}}{\pal t^{2,0}}.
$$
We are to emphasize that, these recursion relations hold true {\it for an arbitrary solution} of extended Toda hierarchy if one defines
the ``correlators'' by the equation (\ref{w-corr}) with the function ${\cal F}$ corresponding to the logarithm of the
tau function of this solution\footnote{In the literature
sometimes these recursion relations together with the Toda equations (\ref{flow2}) are called Toda conjecture.}.
The needed solution is specified by (\ref{tau-rec1}), (\ref{tau-rec2}) together with the string equation.
In this case the recursion relations describe the topology of the forgetting map \cite{KM}
$$
\bar M_{g,n}(CP^1) \to \bar M_{g,n-1}(CP^1).
$$

Due to the discussion of the last Section, we can equally state Theorem \ref{t-c} as follows.
The free energy (\ref{free})
is the logarithm of a particular tau function of the extended NLS hierarchy (\ref{KB-Lax}).

The proof of Theorem \ref{t-c} at the genus one approximation can be found in \cite{D1, DZ1, Egu1, Egu5, Egu6, Z},
see also important results concerning such relations in \cite{getzler, oko2, oko3}.
The crucial point in proving the validity of this conjecture in full genera is the Givental's result on the
Virasoro constraints for $CP^1$ \cite{givental1, givental2}. Probably, one can derive our Toda conjecture from the results of Okounkov and
Pandharipande \cite{oko2, oko3} using the arguments of Getzler's paper \cite{getzler} along with the Givental's result. From our point of view the most natural way of proving the Conjecture
is that to use the properties of the Virasoro symmetries
of the extended Toda hierarchy and the uniqueness of solution of the loop equation \cite{DZ2, DZ3}.
We will publish the details of the proof in a separate paper.
\vskip 0.5truecm
\noindent{\bf Acknowledgments.} The researches of B.D. were
partially supported by Italian Ministry of Education research grant ``Geometry
of Integrable Systems''. The researches of Y.Z. were partially supported
by the Chinese National Science Fund for Distinguished Young Scholars grant
No.10025101 and the Special Funds of Chinese Major Basic Research Project
``Nonlinear Sciences''.

\end{document}